# IONORT: IONOsphere Ray-Tracing

**Programma di ray-tracing nel magnetoplasma ionosferico**

Cesidio Bianchi, Alessandro Settimi*, Adriano Azzarone

*Istituto Nazionale di Geofisica e Vulcanologia (INGV) –*
*Sezione Roma 2 -*
*via di Vigna Murata 605, I-00143 Roma, Italia*

*Corresponding author: Dr. Ing. Alessandro Settimi
Tel: +39-0651860719
Fax: +39-0651860397
Email: alessandro.settimi@ingv.it

**INDICE**

## Premessa

Il pacchetto applicativo "IONORT" per il calcolo del ray-tracing può essere utilizzato dagli utenti che impiegano il sistema operativo Windows. È un programma la cui interfaccia grafica con l'utente è realizzata in MATLAB. In realtà, il programma lancia un eseguibile che integra il sistema d'equazioni differenziali scritto in linguaggio Fortran e ne importa l'output nel programma MATLAB, il quale genera i grafici e altre informazioni sul raggio.

A completamento di questa premessa va detto che questo pacchetto, nella sua parte computazionale, è figlio di un programma di Jones e Stephenson del 1975, dal titolo "*A versatile three-dimensional ray-tracing computer program for radio waves in the ionosphere*", il quale a sua volta si rifaceva principalmente a un programma di ray-tracing di Dudziak (1961) e di altri ricercatori quali Croft and Gregory (1963), ecc.. Pertanto, come tutti i recenti programmi di ray- tracing, questo è un programma fatto di programmi e non si può non menzionare qui la prima applicazione numerica di ray-tracing di Haeselgrove (1955). Attualmente questi programmi sono stati ottimizzati e adattati alle applicazioni dei radar oltre l'orizzonte - Over The Horizon, OTH – [Coleman, 1998][Nickish, 2008] sfruttando le potenzialità di potenti computer e periferiche per la presentazione e l'utilizzo *real-time* nel problema delle *coordinate registration* CR.

In ultimo, si precisa che tutti i parametri di input, output e le modalità d'uso del pacchetto applicativo sviluppato saranno forniti nel manuale utente allegato al CD.

## 1. Introduzione

Il ray-tracing è una tecnica numerica che consente di calcolare il percorso seguito dal raggio d'onda nel magnetoplasma ionosferico. Tramite questa tecnica si determinano i punti dello spazio toccati dal vettore d'onda **k** durante la propagazione. In un mezzo continuo e isotropo, omogeneo o disomogeneo, com'è noto, la traiettoria del raggio (o raggio d'onda), in accordo con la legge di rifrazione, può non seguire un percorso rettilineo ma non lascia mai il piano dove giace in origine il vettore **k**. Nella pratica utilizzante la geometria sferica tale piano è il piano azimutale. Generalmente, in un mezzo anisotropo quale è il magnetoplasma ionosferico, il raggio d'onda devia considerevolmente dal piano azimutale non appena entra nel magnetoplasma e, fintanto che l'onda interagisce con il mezzo, subisce questa deviazione che termina al cessare dell'interazione. Questo porta a deviazioni di svariati km rispetto al piano azimutale d'origine del vettore d'onda. L'entità della deviazione dipende soprattutto dalla direzione del campo magnetico rispetto alla direzione del vettore d'onda. Perciò la complessità di questi fenomeni comporta la formulazione di una teoria che di cui si parlerà brevemente. Si tratta quindi di descrivere la propagazione del raggio in un mezzo continuo disomogeneo e anisotropo con gradienti non troppo elevati. In uno spazio di coordinate e momenti quadridimensionale con il formalismo di Hamilton, si possono scrivere almeno 8 equazioni differenziali che integrate forniscono il percorso del raggio d'onda, il cammino di fase e di gruppo e, in caso di spazio complesso, anche l'assorbimento subito dall'onda.

La teoria conduce alla formulazione delle equazioni canoniche o Hamiltoniane relative al raggio d'onda. Tale teoria si trova descritta in numerosi testi tra cui: Budden (1985), Kelso (1964), vari altri non meno importanti autori e anche riproposta da Bianchi (2009) in un rapporto tecnico dell'INGV.

Questo lavoro prende le mosse soprattutto dal programma di Jones e Stephenson, molto diffuso nella comunità scientifica, che si interessa di radio propagazione per via ionosferica. Il programma è scritto in linguaggio FORTRAN 77, per un *mainframe* CDC-3800. Il codice in sé, oltre ad essere molto elegante, è estremamente efficiente e costituisce la base di numerosi programmi ora in uso soprattutto nell'applicazioni di CR dei radar OTH. L'input e l'output di questo programma necessita di periferiche non più in uso da vari decenni e non esistono compilatori che accettino le istruzioni scritte per quel tipo di *mainframe*. Per questa ragione il nucleo del programma in grado di eseguire le integrazioni numeriche, dopo le necessarie modifiche, è stato passato a un moderno compilatore sotto il sistema operativo Windows e l'eseguibile è stato importato in un programma MATLAB. Quindi, tutte le operazioni di input e output vengono trattate dal moderno programma MATLAB che esegue il programma Fortran e ne importa l'output. Questo conferisce una grande versatilità all'intero pacchetto applicativo con presentazioni in due dimensioni (2D) e tre dimensioni (3D) geo-referenziate su mappe reali.

## 2. Il problema generale del ray-tracing ionosferico

Il calcolo del ray-tracing in ionosfera, una volta nota la densità elettronica N in tutti i punti che le radio onda investe, è un calcolo perfettamente deterministico. L'accuratezza del calcolo, se si trascura la discretizzazione dello spazio e del tempo dovuta al processo d'integrazione, dipende solamente dall'accuratezza con cui si conosce la densità elettronica N e dal campo magnetico terrestre. Se siamo interessati all'assorbimento dell'onda, essa dipende anche dalla densità dell'atmosfera neutra. Immaginiamo di avere un plasma senza campo magnetico, come in figura 1: una volta conosciuta la grandezza che meglio lo rappresenta (la densità elettronica N), forniti alcuni parametri iniziali quali latitudine, longitudine e altezza e stabilita la direzione dell'onda (direzione del vettore **k**), il calcolo di ray-tracing permette di determinare i punti toccati dall'apice del vettore **k**, cioè il percorso esatto del raggio d'onda. Dato che ogni punto della superficie terrestre ha un profilo verticale di densità elettronica, profili rappresentati dalla matrice 3D in figura, il raggio d'onda investe tali profili a differenti altezze con un'interazione che riguarda solo il profilo a quella quota.

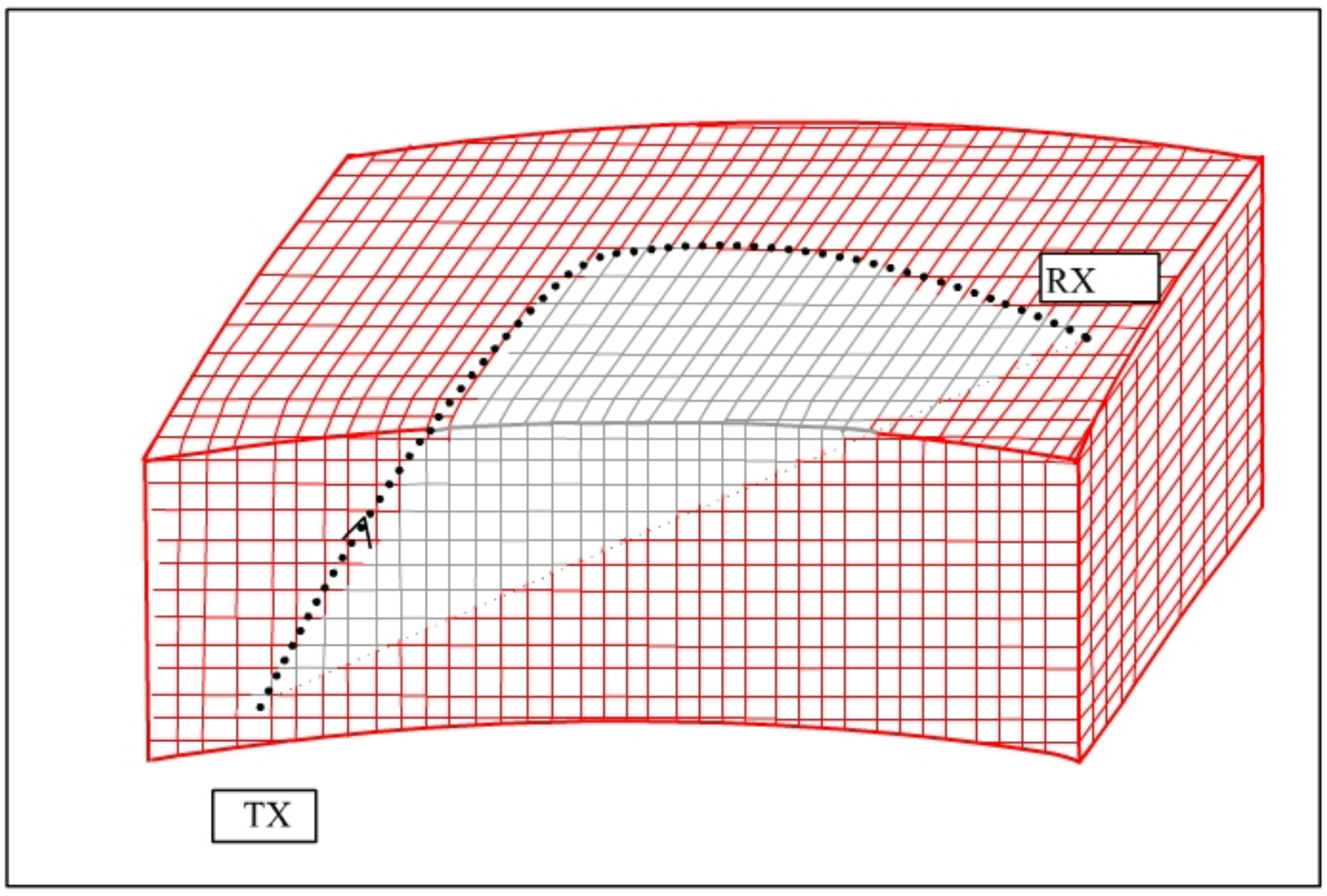

**Figura 1.** Percorso del raggio d'onda in ionosfera dal trasmettitore (TX) al ricevitore (RX). In evidenza, i profili di densità elettronica interessati lungo il percorso del raggio d'onda nella matrice ionosferica 3D.

La trattazione sin qui svolta vale nell'ipotesi esemplificativa adottata di un mezzo isotropo che si verifica solo in particolari situazioni operative. In un mezzo isotropo, il raggio giace sul piano azimutale evidenziato in figura 1. e non vi è ragione fisica per la quale il raggio abbandoni questo piano d'origine. Nella generalità dei casi, il mezzo deve essere considerato anisotropo per effetto del campo magnetico terrestre e pertanto la geometria di propagazione dipende dalla direzione. La precedente figura viene così a modificarsi come indicato nella figura 2.

A titolo d'esempio, se il percorso della radio onda avviene con un angolo azimutale $\alpha$ rispetto al meridiano magnetico, esso può deviare di un angolo $\Delta\alpha$ che si può determinare con un ray-tracing completo. Il raggio d'onda può deviare di un $\Delta\alpha$ fino a 3° gradi rispetto al piano azimutale di partenza. La deviazione dal piano azimutale avviene soltanto lungo il tragitto in ionosfera (linea tratteggiata).

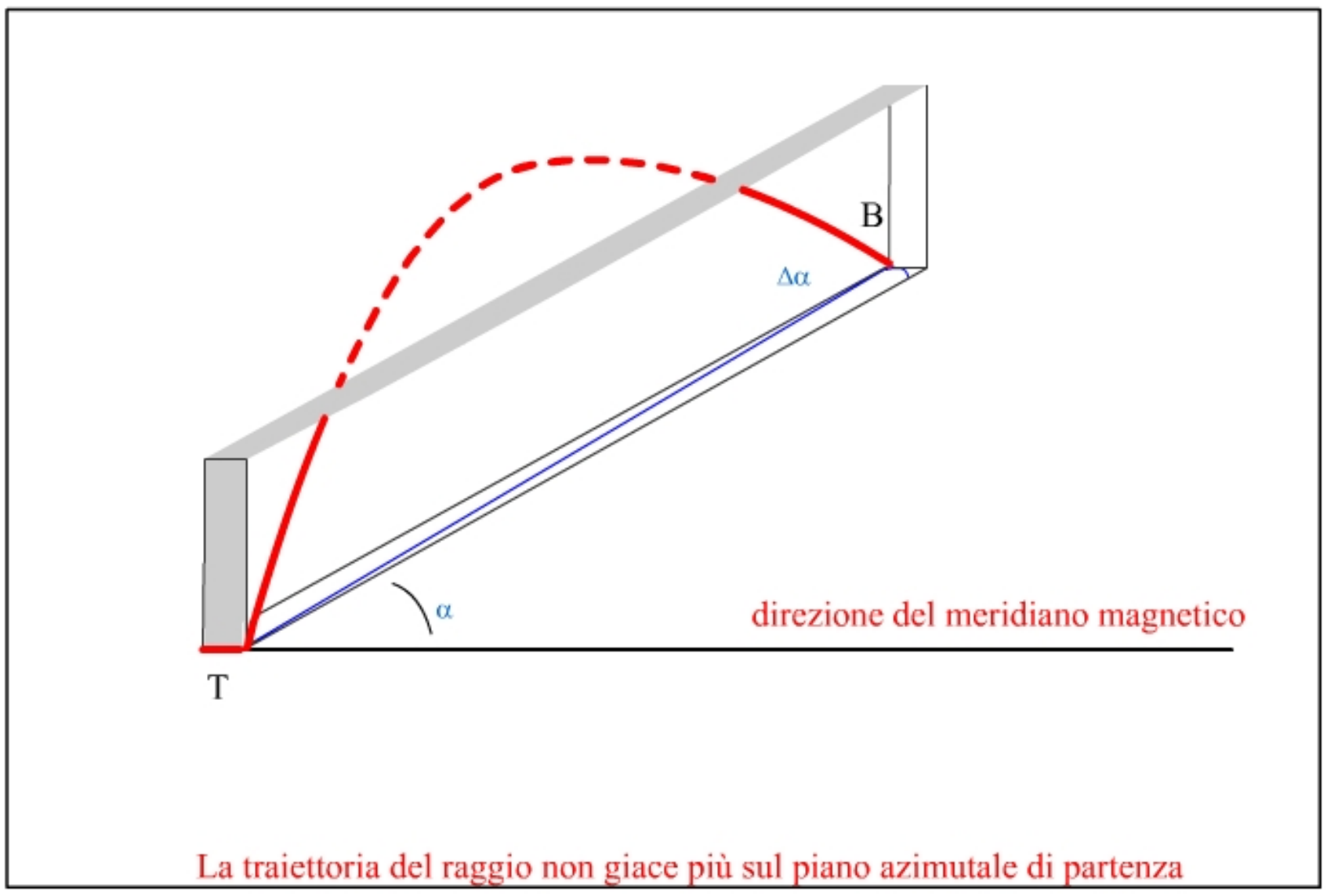

**Figura 2.** (Tratta da Kelso 1964) Percorso su un piano azimutale non coincidente con il meridiano magnetico.

Per questa ragione il problema generale del ray-tracing diventa estremamente complesso sul piano teorico (vedi appendice A) e altrettanto complicato sul piano puramente computazionale. In effetti siamo costretti a usare un Hamiltoniana che, in ultima analisi, dipende dall'indice di rifrazione dato dalla formula di Appleton-Hartree riportata in appendice. Vi sono due ordini di difficoltà. Il primo è legato al doppio segno che appare nella formula che comporta la propagazione di due raggi, cosiddetti ordinario e straordinario. Ciascuno di questi raggi si propaga con un suo proprio valore dell'indice di rifrazione, come avviene nei cristalli anisotropi [Fowles, 1989]. Ma questa è una difficoltà che presenta più aspetti concettuali che computazionali, comunque superabili. Il secondo ordine di difficoltà deriva dal fatto che le funzioni multi-valore, che scaturiscono dall'indice di rifrazione avente doppi segni e radice ed anche dalle particolari condizioni geometriche (es. divisioni per $\cos\beta$, con $\beta$ che tende a zero), poco si prestano per essere tradotte in linguaggio matematico non analitico, quali sono i linguaggi di programmazione.

## 3. Descrizione del programma

Il software di ray-tracing in questione ha i parametri di input e output gestiti da un programma scritto in MATLAB. Quindi l'utente finale vede solo un'interfaccia che accetta dati in ingresso relativi alla posizione del trasmettitore (latitudine, longitudine, quota), angoli di azimut e d'elevazione, frequenza e altri parametri computazionali sul ray-tracing. Al nucleo, che contiene l'algoritmo principale di calcolo, vengono fornite tutte le informazioni di input tramite un vettore reale.

In uscita, il software utilizza tutte le potenzialità di un programma ad alto livello per la stampa di immagini. Le figure 2D e 3D costituiscono un output affidabile dell'interfaccia grafica in quanto sono la rappresentazione grafica di un file numerico prodotto dal nucleo di un programma, a sua volta basato su una teoria del ray tracing, che sono stati ampiamente verificati e quindi non affetti da errori.

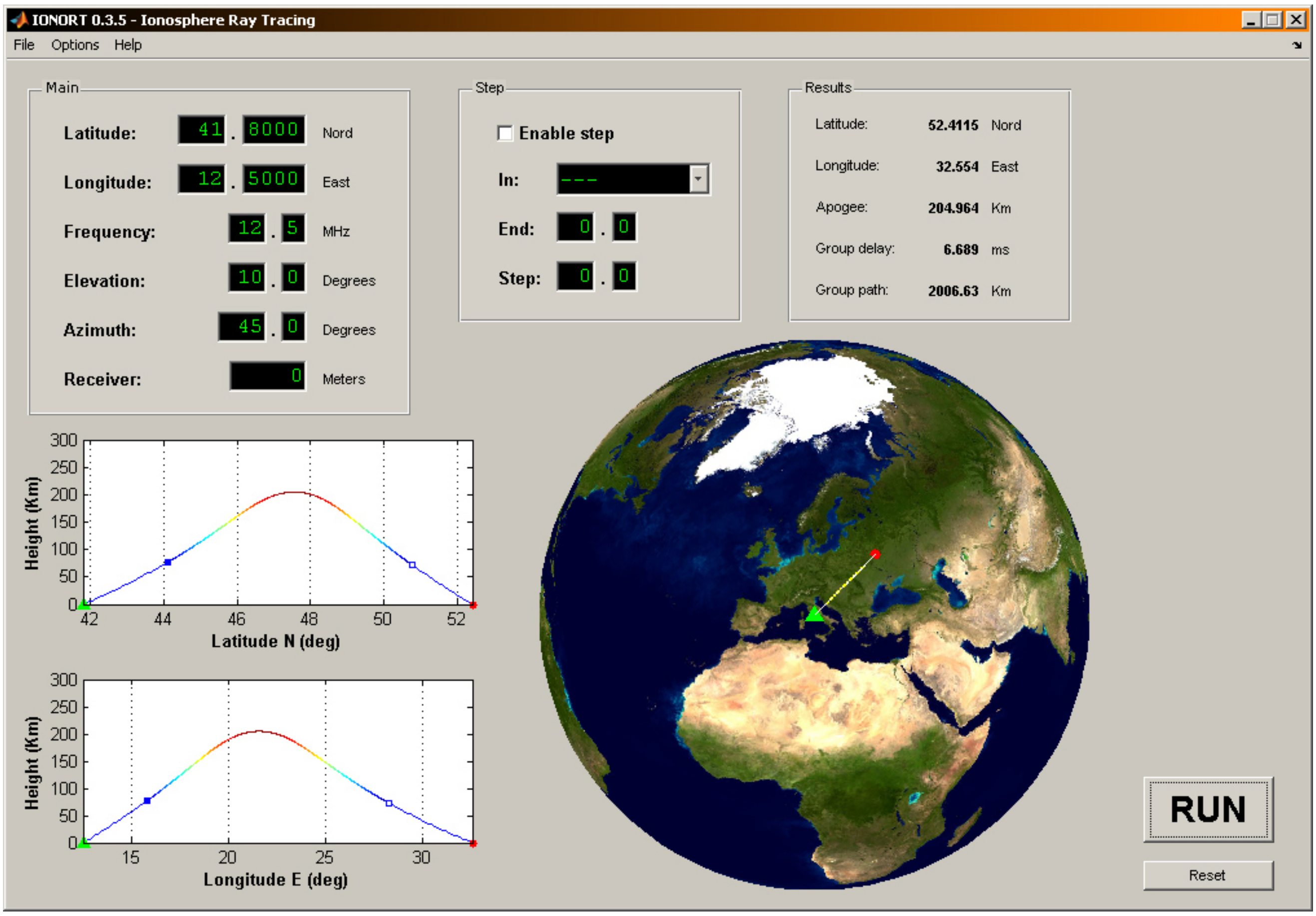

**Figura 3.** Stadio di sviluppo attuale dell'interfaccia grafica del programma IONORT

La parte del programma che esegue tutti gli algoritmi di calcolo relativi al ray-tracing è in un file eseguibile generato dalla compilazione del codice Fortran. Questo pacchetto coniuga l'estrema velocità computazionale di un eseguibile Fortran con la praticità di un'interfaccia grafica consentita da MATLAB.

| INPUT |
|---|
| 1. Lettura MATLAB variabili default file .ini<br>2. Visualizzazione interfaccia<br>3. L'utente modifica i valori di default e avvia il calcolo<br>4. Generazione file di testo con le variabili per il Fortran |
| **ESEGUIBILE** |
| Algoritmo di calcolo di ray-tracing in Fortran costituito principalmente da sub-routine per il calcolo dei modelli e l'integrazioni delle equazioni differenziali |
| **OUTPUT** |
| 1. Stampa delle coordinate a terminale<br>2. Lettura MATLAB del terminale<br>3. Generazione di grafici 2D e 3D e stampa dei risultati significativi dell'elaborazione nell'interfaccia<br>4. Salvataggio sul computer dei grafici e dei risultati |

**Tabella 1.** Schema di composizione e funzione delle tre sezioni del programma.

Il programma è in grado di eseguire il calcolo anche su intervalli di valori in frequenza, oppure in angolo d'elevazione o di azimut. MATLAB lancia l'eseguibile ciclicamente, variando ogni volta il valore del parametro, secondo il passo deciso dall'utente. Il risultato di tale operazione è rappresentato nei due esempi delle figure 4 e 5. Le immagini sono in 2D e quindi non occorre georeferenziazione.

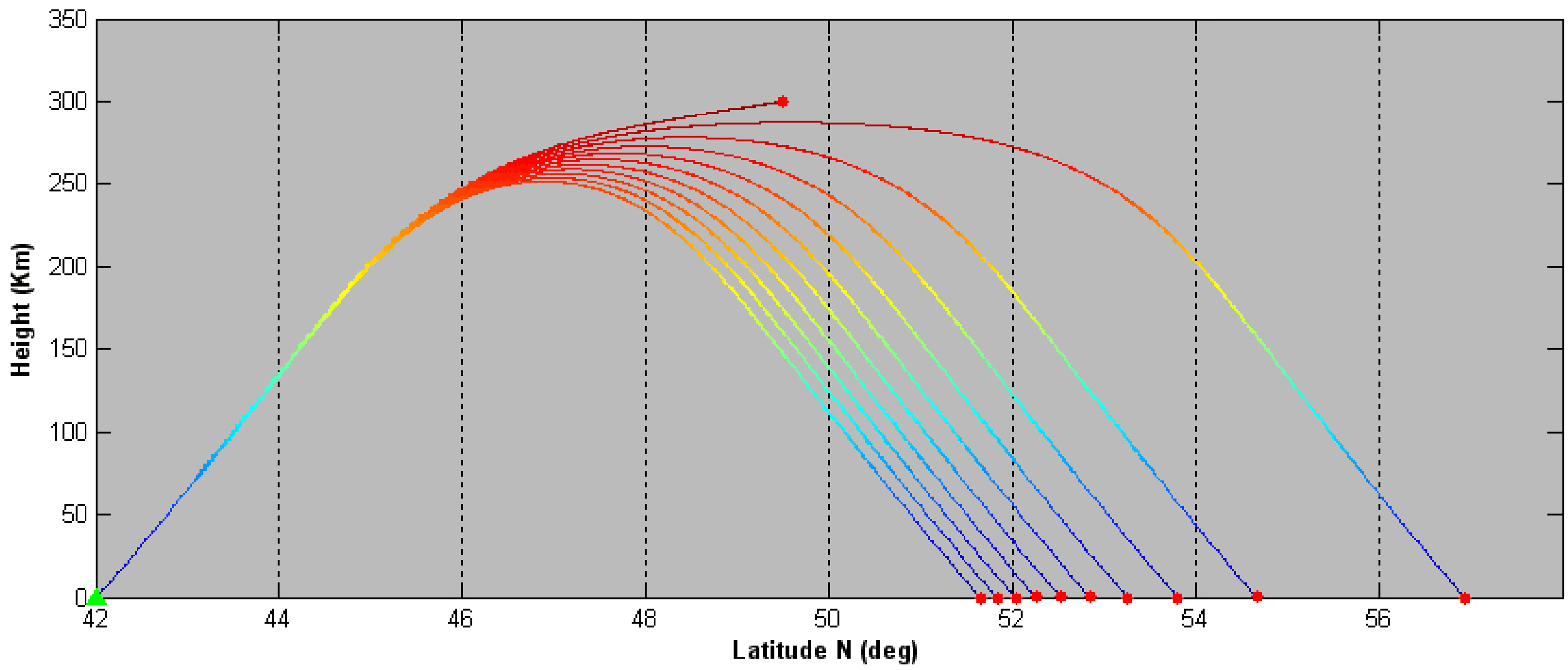

**Figura 4.** Esempio: variazione in frequenza da 10.0 a 11.0 MHz con passo 0.1, Elevazione 30° e Azimut 0°, Coordinate 42° N - 12.5° E. Da sinistra a destra, raggi d'onda a frequenze crescenti. Si noti come alla frequenza più alta il raggio penetra la ionosfera (l'asse delle ordinate è amplificato di un fattore 2 rispetto l'ascissa).

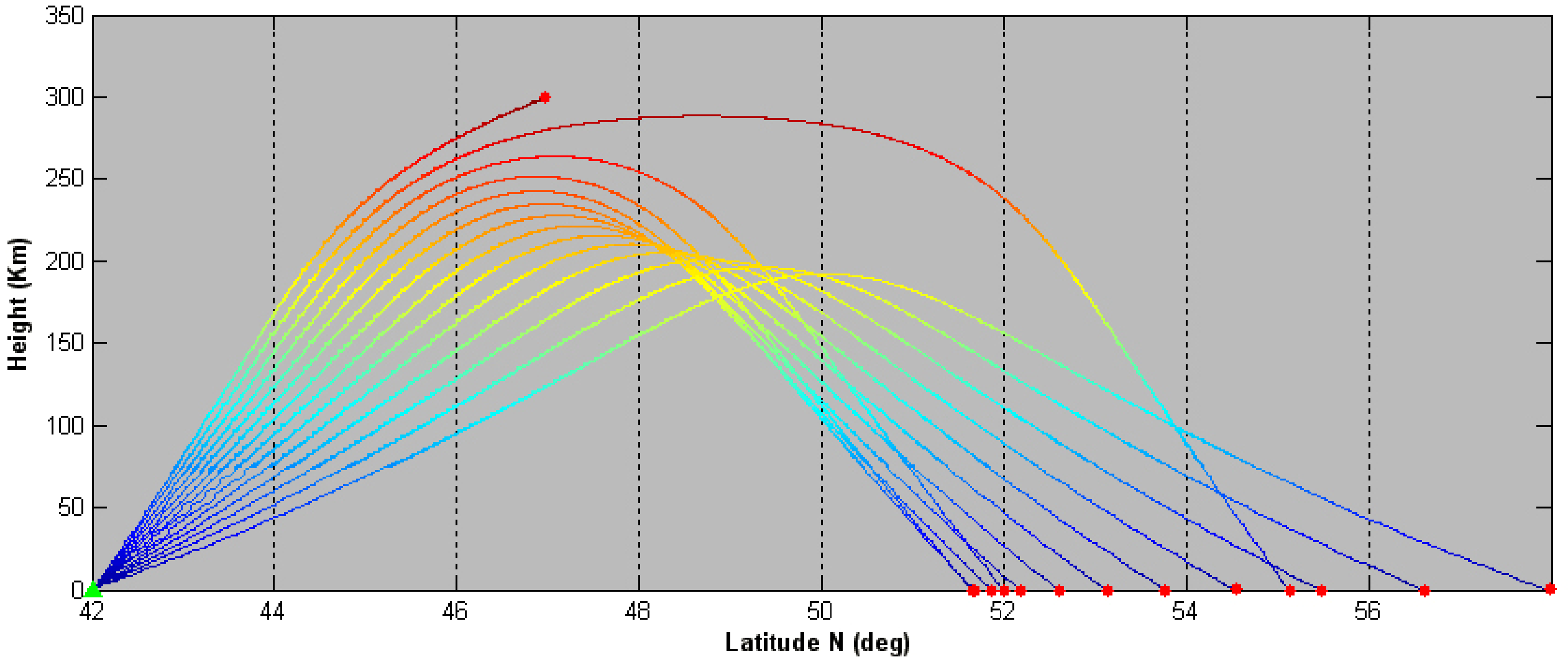

**Figura 5.** Esempio: variazione in elevazione da 10 a 36° con passo 2, Frequenza 10 MHz, Azimuth 0°, Coordinate 42° N – 12.5° E. Da sinistra a destra, raggi d'onda con elevazione decrescente. Si noti che intorno all'elevazione limite di 29° il raggio permane a quote alte in ionosfera (angolo critico di riflessione) e arriva a terra all'incirca alla stessa distanza raggiunta dal raggio con elevazione compresa tra 14° e 16°( l'asse delle ordinate è amplificato di un fattore 2 rispetto l'ascissa.).

L'applicativo consente di usufruire di funzioni di grafica avanzate, permettendo una rappresentazione 3D realistica con una precisa geo-referenziazione come negli esempi di figura 6 e 7.

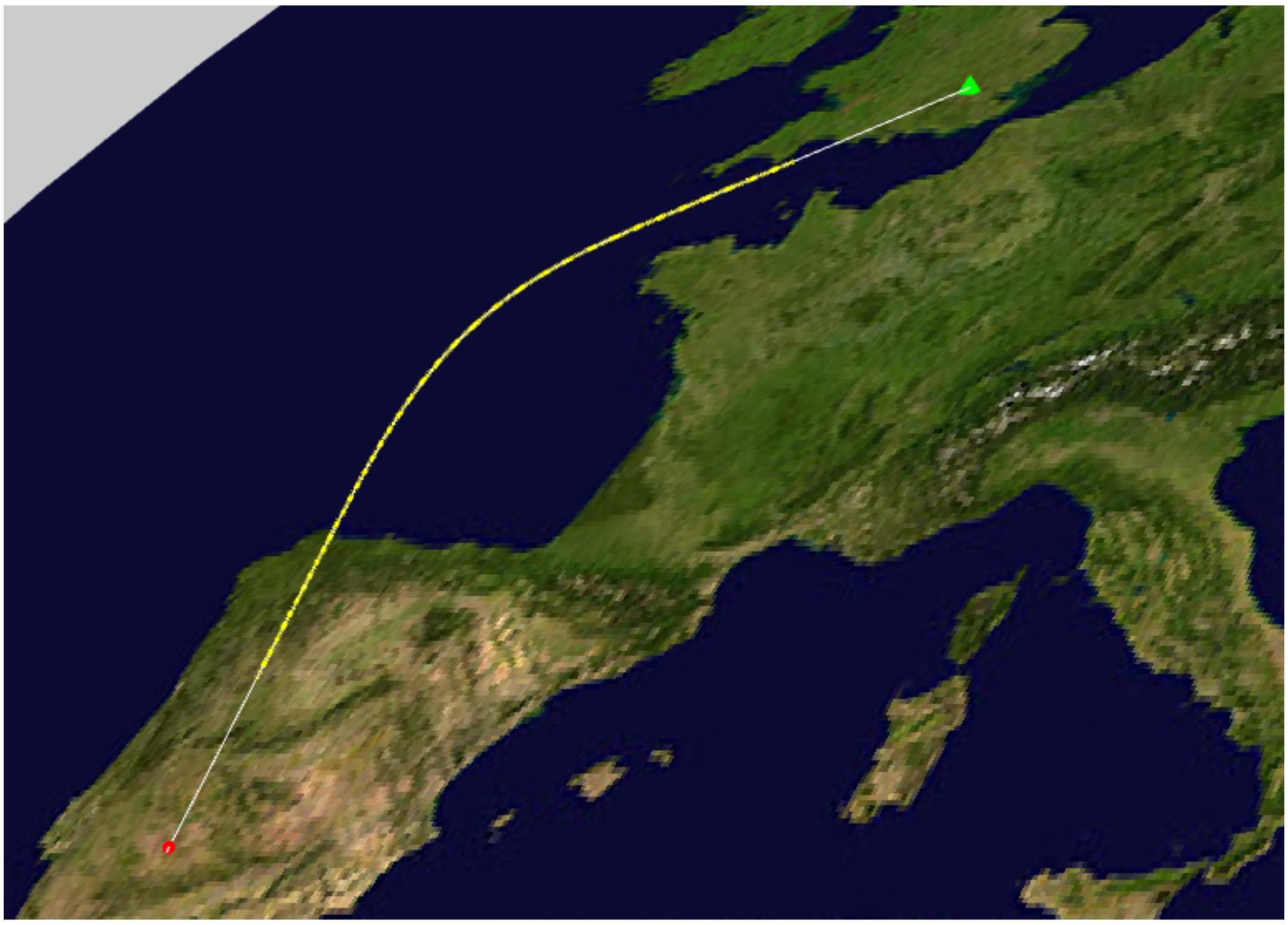

**Figura 6.** Visualizzazione in 3D di un raggio partito dall'Osservatorio di Greenwich (51.4772° - N 0° E) a Frequenza 11 MHz, Elevazione 15° e Azimut 200°. La parte evidenziata in giallo rappresenta il cammino del raggio nella ionosfera (errore relativo sulle coordinate geografiche di $10^{-4}$).

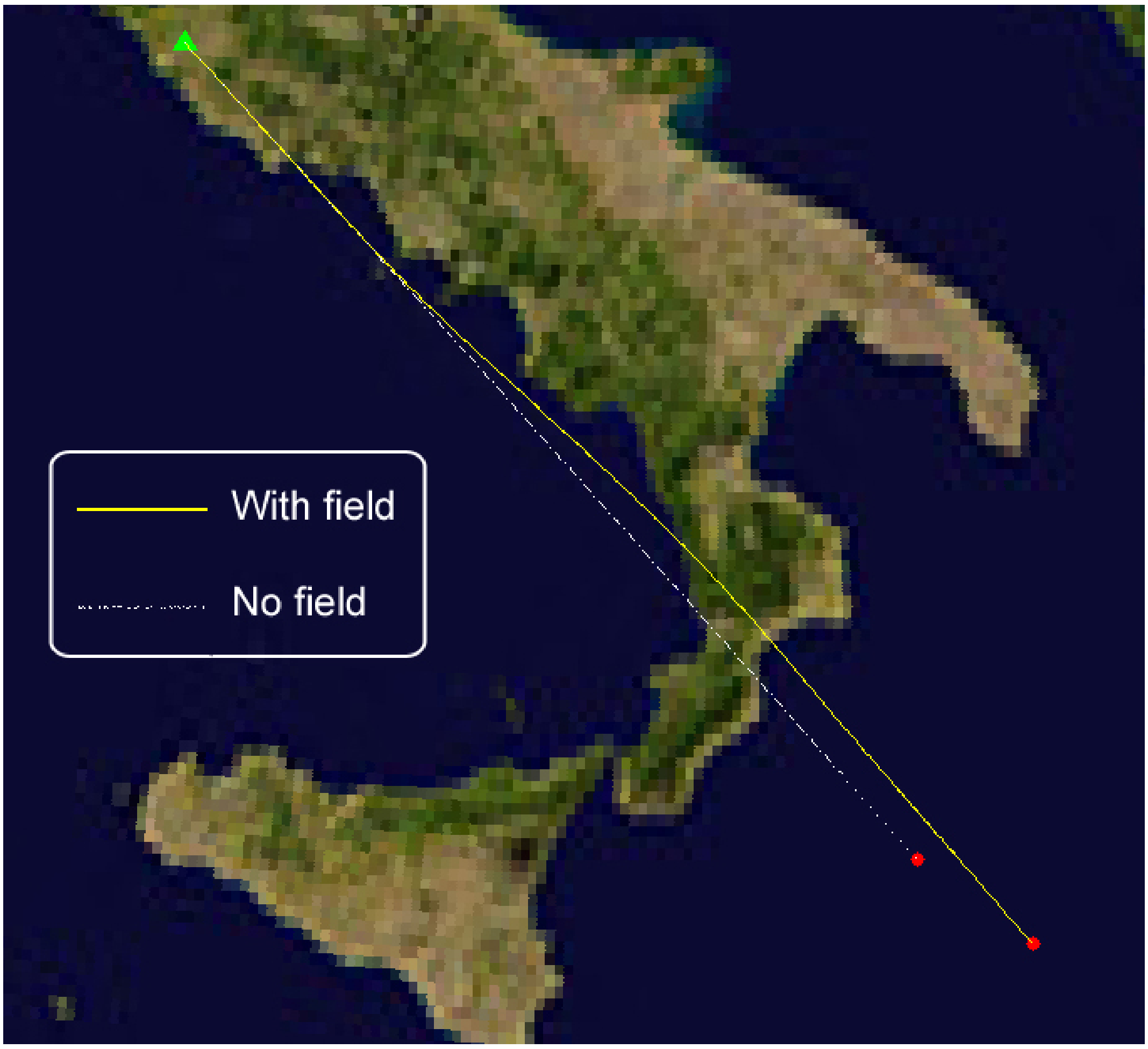

**Figura 7.** Percorso di raggio con e senza campo magnetico. Coordinate 41.8° N - 12.5° E, Frequenza 8 MHz, Elevazione 45°, Azimut 135° (errore relativo sulle coordinate geografiche di $10^{-4}$).

La parte relativa alle variabili in ingresso merita un'attenzione particolare in quanto è l'unico modo per un utente di interagire con il programma Fortran. Tutte queste variabili sono contenute in un file di testo che viene letto dal programma come un vettore W di 400 componenti di cui diamo le specifiche. In tabella 2 sono riportate le locazioni da 1 a 25 delle variabili in ingresso riguardanti: le coordinate (trasmettitore, ricevitore e polo magnetico), frequenza, azimut, angolo di elevazione, tipo di raggio (ordinario o straordinario), numero di salti in ionosfera e massimo numero di passi nella subroutine TRACE.

In tabella 3 sono riportate le componenti del vettore W riguardanti la subroutine d'integrazione RKAM (Runge-Kutta, Adams-Moulton). Qui l'utente sceglie tra i due algoritmi: quello deterministico Runge-Kutta e quello adattivo del tipo predictor-corrector di Adams-Moulton. Oltre alle equazioni descritte in appendice, il programma integra anche le altre equazioni relative al percorso di fase, assorbimento Doppler e percorso geometrico effettivo (lungo la coordinata curvilinea del raggio). Vi sono poi tutta una serie di parametri relativi al passo d'integrazione e controllo sugli errori.

| Posizione | Variabile | Descrizione | Valori |
|---|---|---|---|
| W1 | **RAY** | Raggio o/ex | 1. = ordinario<br>2. = straordinario |
| W2 | **EARTH** | Raggio della Terra | 6371 km |
| W3 | **XMTRH** | Altezza TX | km |
| W4 | **TLAT** | Latitudine TX | rad |
| W5 | **TLON** | Longitudine TX | rad |
| W6 | **F** | Frequenza | MHz |
| W7 | **FBEG** | Frequenza iniziale | MHz |
| W8 | **FEND** | Frequenza finale | MHz |
| W9 | **FSTEP** | Frequenza -passo | MHz |
| W10 | **AZI** | Azimut | rad |
| W11 | **AZBEG** | Azimut iniziale | rad |
| W12 | **AZEND** | Azimut finale | rad |
| W13 | **AZSTEP** | Azimut - passo | rad |
| W14 | **BETA** | Angolo elevazione | rad |
| W15 | **ELBEG** | Elevazione iniziale | rad |
| W16 | **ELEND** | Elevazione finale | rad |
| W17 | **ELSTEP** | Passo in elevazione | rad |
| W20 | **RCVRH** | Altezza RX | km |
| W21 | **ONLY** | Numero di passi dopo la penetrazione | reale |
| W22 | **HOP** | Numeri di salti | reale |
| W23 | **MAXSTP** | Massimo numero di passi | reale |
| W24 | **PLAT** | Latitudine polo N | rad |
| W25 | **PLON** | Longitudine polo N | rad |

**Tabella 2.** Variabili di inizializzazione

| Posizione | Variabile | Descrizione | Valori |
|---|---|---|---|
| W41 | **INTYP** | Algoritmo d'integrazione | 1. = Runge-Kutta<br>2. = Adams-Moulton (senza controllo sull'errore)<br>3. = Adams-Moulton (contr. sull'errore rel.)<br>4. = Adams-Moulton (contr. sull'errore ass.) |
| W42 | **MAXERR** | Passo - Errore massimo | $10^{-4}$ |
| W43 | **ERATIO** | Rapporto errore max/min | 20 |
| W44 | **STEP1** | Cammino di gruppo - Passo | 0.05 km |
| W45 | **STPMAX** | Passo massimo | 13.5 km |
| W46 | **STPMIN** | Passo minimo | $10^{-8}$ km |
| W47 | **FACTR** | Fattore agente sul passo | 0.5 |
| W57 | | Percorso di fase | 1. = integrazione<br>2. = integrazione e stampa |
| W58 | | Assorbimento | = 1. oppure = 2. |
| W59 | | Spostamento di Doppler | = 1. oppure = 2. |
| W60 | | Lunghezza del percorso | = 1. oppure = 2. |

**Tabella 3.** Parametri per l'algoritmo d'integrazione e dati di default, inizializzati dal programma IONORT.

Le componenti da 100 a 149 del vettore W riservano le locazioni dei parametri relativi ai modelli di densità elettronica in uso nel programma. Ad esempio, nella tabella 4 è stato inserito un modello di Chapman per la densità elettronica impostato nella seguente maniera:

| Posizione | Variabile | Descrizione | Valori |
|---|---|---|---|
| W101 | FC | Frequenza critica all'equatore | 6.5 MHz |
| W102 | HM | Altezza della massima densità elettronica all'equatore | 300 km |
| W103 | SH | Altezza di scala | 62 km |
| W104 | ALPHA | Strato di Chapman alpha o beta | 0.5 ÷ 1.0 |
| W105 | A | Ampiezza della variazione periodica di $(FC)^2$ con la latitudine | 0 |
| W106 | B | Periodo di variazione di $(FC)^2$ con la latitudine | 0 |
| W107 | C | Coefficiente della variazione lineare di $(FC)^2$ con la latitudine | 0 |
| W108 | E | Inclinazione dello strato di Chapman | 0 |

**Tabella 4**. Parametri del modello di Chapman di densità elettronica, inseriti dall'utente nel file testo del vettore W.

In figura 8 è riportato il diagramma a blocchi del nucleo centrale del programma eseguibile con le principali subroutine.

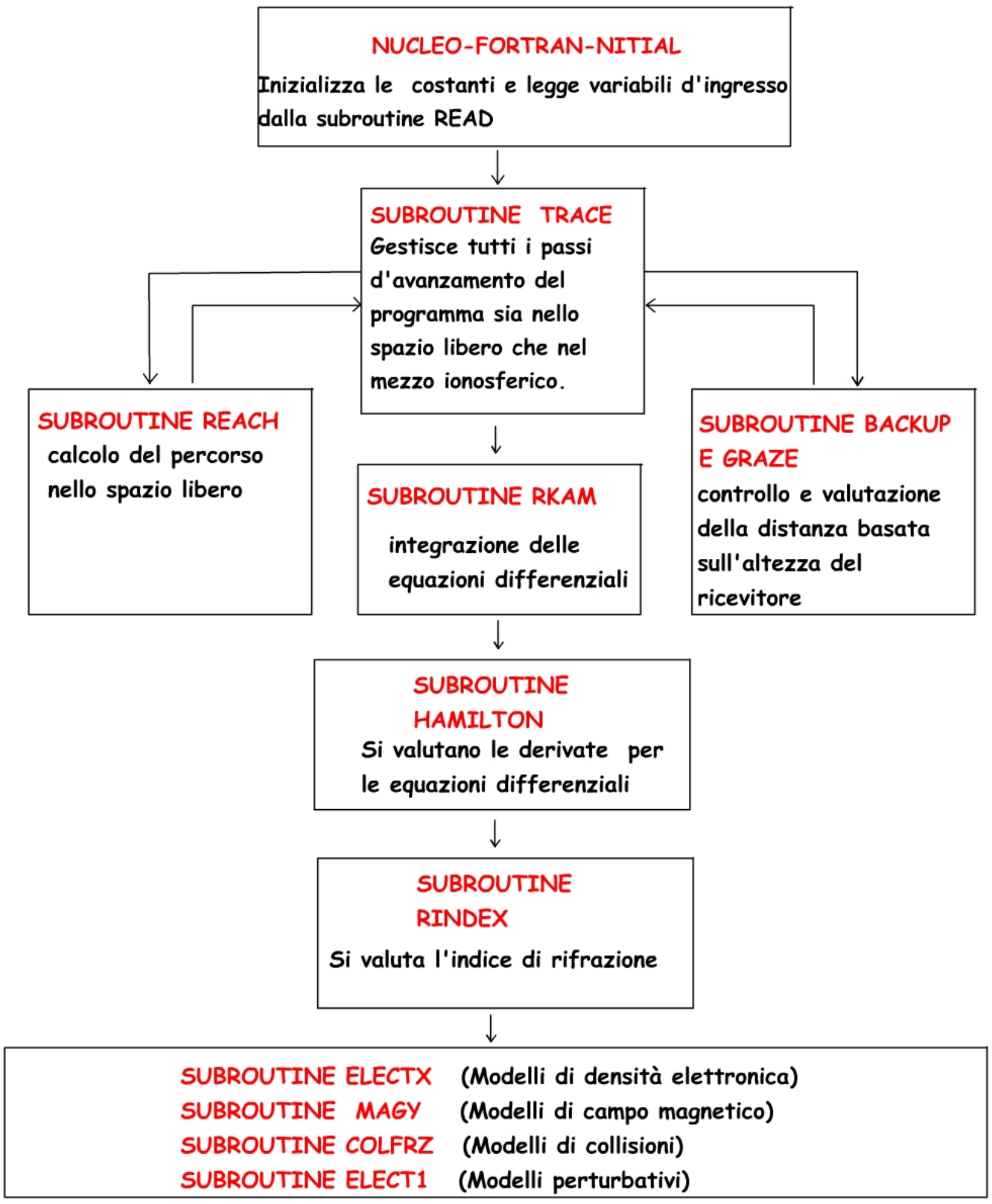

**Figura 8.** Nucleo del programma di calcolo

## 4. Appendice

In questa appendice, si svolge un importante richiamo alla teoria di ray tracing sottostante il funzionamento del programma IONORT, per chiarire la sua implementazione nel nucleo del programma realizzato da Jones e Stevenson (1975). In uno spazio di coordinate generalizzate a quattro dimensioni, $p_i$ le tre coordinate spaziali con l'aggiunta di $t$ e $q_i$ i tre momenti generalizzati con l'aggiunta di $\omega$, si può scrivere la coppia di equazioni (Weinberg 1962)

$$\frac{\partial H(q_i, p_i)}{\partial p_i} = \frac{dq_i}{d\tau}, \tag{1a}$$

$$\frac{\partial H(q_i, p_i)}{\partial q_i} = -\frac{dp_i}{d\tau}, \tag{1b}$$

dove $H(t, x, y, z, k_x, k_y, k_z, \omega)$ è l'Hamiltoniana del mezzo e $\tau$ variabile indipendente di $H$. Da queste scaturiscono 8 equazioni che possono essere integrate per avere i valori delle coordinate e del tempo dell'evoluzione del raggio lungo il percorso dell'onda, nonché i momenti generalizzati che qui sono visti come le componenti del vettore d'onda e la pulsazione $\omega$.

Le equazioni (1) sono chiamate equazioni canoniche del raggio e in questo contesto assumono la forma familiare della formulazione Hamiltoniana. Esse sono molto generali poiché includono il caso in cui il mezzo è dispersivo, cioè dipendente esplicitamente dalla frequenza, e variante nel tempo, cioè dipendente esplicitamente dal tempo. In generale, $H$ è anche una grandezza complessa cioè composta da una parte reale e una parte immaginaria

$$H(t, x, y, z, k_x, k_y, k_z, \omega) = \frac{c}{\omega} \frac{\sqrt{k_x^2 + k_y^2 + k_z^2}}{n(t, x, y, z, k_x, k_y, k_z, \omega)} = 1, \tag{2}$$

essendo $\frac{\omega}{c} n = k$ con $n=(\mu+j\chi)$ indice di rifrazione complesso. Ciò posto, questa potrebbe essere una possibile Hamiltoniana che è una costante lungo il tragitto del raggio d'onda. Con alcune posizioni algebriche, si può prendere solo la parte reale di questa o la parte immaginaria. Quando la lunghezza d'onda è relativamente grande rispetto alla scala spaziale in cui l'onda può essere significativamente assorbita, la parte reale di $H$ consente da sola di ottenere un ray-tracing del mezzo. Nel lavoro di Jones and Stephenson (1975) $H$ è posta come nella seguente forma

$$H(t, x, y, z, k_x, k_y, k_z, \omega) = \mathrm{Re}\left\{ \frac{1}{2} [\frac{c^2}{\omega^2}(k_x^2 + k_y^2 + k_z^2) - n^2] \right\}, \tag{3}$$

oppure in coordinate sferiche si può avere:

$$H(t, r, \theta, \phi, k_r, k_\theta, k_\phi, \omega) = \mathrm{Re}\left\{ \frac{1}{2} [\frac{c^2}{\omega^2}(k_r^2 + k_\theta^2 + k_\varphi^2) - n^2] \right\}. \tag{4}$$

Si possono scrivere molte varianti simili a quest'ultima equazione, ma tutte si basano sul fatto che $H$ è una costante di propagazione del raggio. Poiché qui siamo interessati a un problema di ray-tracing, per amore di semplicità, possiamo continuare la trattazione considerando per il momento solo la parte reale dell'indice di rifrazione $\mu$.
Esplicitamente le equazioni che scaturiscono dalle (1) sono

$$\frac{dx}{d\tau}=\frac{\partial H}{\partial k_x}, \qquad (5.a)$$

$$\frac{dy}{d\tau}=\frac{\partial H}{\partial k_y}, \qquad (5.b)$$

$$\frac{dz}{d\tau}=\frac{\partial H}{\partial k_z}, \qquad (5.c)$$

$$\frac{dt}{d\tau}=-\frac{\partial H}{\partial \omega}, \qquad (5.d)$$

$$\frac{dk_x}{d\tau}=-\frac{\partial H}{\partial x}, \qquad (5.e)$$

$$\frac{dk_y}{d\tau}=-\frac{\partial H}{\partial y}, \qquad (5.f)$$

$$\frac{dk_z}{d\tau}=-\frac{\partial H}{\partial z}, \qquad (5.g)$$

$$\frac{d\omega}{d\tau}=\frac{\partial H}{\partial t}, \qquad (5.h)$$

essendo τ parametro che varia monotonamente lungo il percorso dell'onda. Simili equazioni in coordinate sferiche si possono scrivere come:

$$\frac{dr}{d\tau}=\frac{\partial H}{\partial k_r}, \qquad (6.a)$$

$$\frac{d\theta}{d\tau}=\frac{1}{r}\frac{\partial H}{\partial k_\theta}, \qquad (6.b)$$

$$\frac{d\varphi}{d\tau}=\frac{1}{r\sin\theta}\frac{\partial H}{\partial k_\varphi}, \qquad (6.c)$$

$$\frac{dt}{d\tau}=-\frac{\partial H}{\partial \omega}, \qquad (6.d)$$

$$\frac{dk_r}{d\tau}=-\frac{\partial H}{\partial r}+k_\theta\frac{d\theta}{d\tau}+k_\varphi\sin\theta\frac{d\varphi}{d\tau}, \qquad (6.e)$$

$$\frac{dk_\theta}{d\tau}=\frac{1}{r}\left(-\frac{\partial H}{\partial \theta}-k_\theta\frac{dr}{d\tau}+k_\varphi r\cos\theta\frac{d\varphi}{d\tau}\right), \qquad (6.f)$$

$$\frac{dk_\varphi}{d\tau}=\frac{1}{r\sin\theta}\left(-\frac{\partial H}{\partial \varphi}-k_\varphi\sin\theta\frac{dr}{d\tau}-k_\varphi r\cos\theta\frac{d\theta}{d\tau}\right), \qquad (6.g)$$

$$\frac{d\omega}{d\tau}=\frac{\partial H}{\partial t}. \qquad (6.h)$$

In figura 9 sono rappresentate le coordinate sferiche $r$, $\theta$, $\varphi$, insieme alle componenti del versore d'onda $k_r$, $k_\theta$, $k_\varphi$ , rispetto a un sistema computazionale geocentrico.

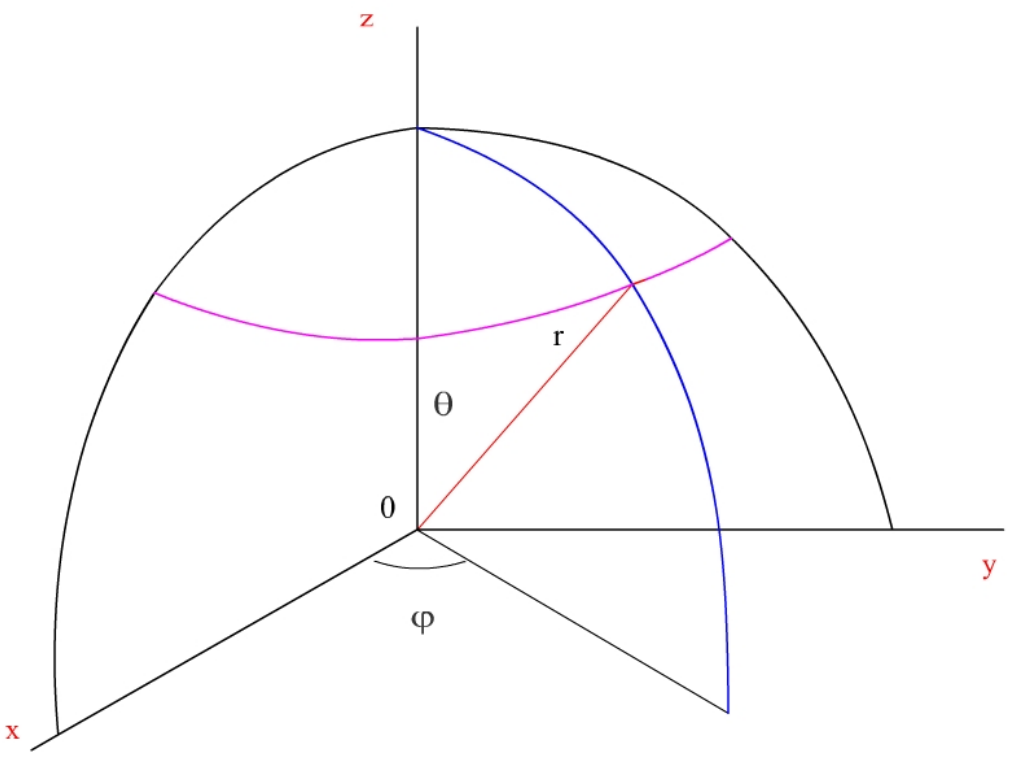

**Figura 9.** Sistema di riferimento cartesiano ($x$, $y$, $z$) e sferico ($r$, $\theta$, $\varphi$)

Il programma di ray-tracing qui riproposto integra le equazioni differenziali (6). Una considerazione va fatta riguardo alla variabile indipendente la cui scelta può facilitare gli algoritmi di calcolo. Ad esempio, si può prendere $\tau=ct=P'$ che è il percorso di gruppo e soddisfa la condizione di monotonia prima detta. Con tale posizione, e applicando la regola della catena nell'esecuzione delle derivate parziali rispetto a $P'$, le (6) diventano:

$$\frac{dr}{dP'}=-\frac{1}{c}\frac{\frac{\partial H}{\partial k_r}}{\frac{\partial H}{\partial \omega}}, \quad (7.a)$$

$$\frac{d\theta}{dP'}=-\frac{1}{rc}\frac{\frac{\partial H}{\partial k_\theta}}{\frac{\partial H}{\partial \omega}}, \quad (7.b)$$

$$\frac{d\varphi}{dP'}=-\frac{1}{rc\sin\theta}\frac{\frac{\partial H}{\partial k_\varphi}}{\frac{\partial H}{\partial \omega}}, \quad (7.c)$$

$$\frac{dt}{d\tau}=-\frac{\partial H}{\partial \omega}, \quad (7.d)$$

$$\frac{dk_r}{dP'}=\frac{1}{c}\frac{\frac{\partial H}{\partial r}}{\frac{\partial H}{\partial \omega}}+k_\theta\frac{d\theta}{dP'}+k_\varphi\sin\theta\frac{d\varphi}{dP'}, \quad (7.e)$$

$$\frac{dk_\theta}{dP'}=\frac{1}{r}\left(\frac{1}{c}\frac{\frac{\partial H}{\partial \theta}}{\frac{\partial H}{\partial \omega}}-k_\theta\frac{dr}{dP'}+k_\varphi r\cos\theta\frac{d\varphi}{dP'}\right), \quad (7.f)$$

$$\frac{dk_\varphi}{dP'}=\frac{1}{r\sin\theta}\left(\frac{1}{c}\frac{\frac{\partial H}{\partial \varphi}}{\frac{\partial H}{\partial \omega}}-k_\varphi\sin\theta\frac{dr}{dP'}-k_\varphi r\cos\theta\frac{d\theta}{dP'}\right), \quad (7.g)$$

$$\frac{d\omega}{d\tau}=\frac{\partial H}{\partial t}. \quad (7.h)$$

Nei gruppi di 8 equazioni si evincono le 8 variabili dipendenti che possono tutte variare lungo il percorso. Ad un certo istante, esse assumono un valore che può essere usato come punto di partenza per successe valutazioni. In altri termini, questo significa che a partire da un certo punto dove l'Hamiltoniana ha un significato fisico, cioè in ionosfera, il primo passo d'integrazione con i valori delle 8 variabili dipendenti è punto di partenza del successivo passo d'integrazione e così via. La scelta della Hamiltoniana (4) porta direttamente alle seguenti equazioni

$$\frac{\partial H}{\partial t} = -n\frac{\partial n}{\partial t}, \tag{8.a}$$

$$\frac{\partial H}{\partial r} = -n\frac{\partial n}{\partial r}, \tag{8.b}$$

$$\frac{\partial H}{\partial \theta} = -n\frac{\partial n}{\partial \theta}, \tag{8.c}$$

$$\frac{\partial H}{\partial \varphi} = -n\frac{\partial n}{\partial \varphi}, \tag{8.d}$$

$$\frac{\partial H}{\partial \omega} = -n\frac{n'}{\omega}, \tag{8.e}$$

$$\frac{\partial H}{\partial k_r} = \frac{c^2}{\omega^2}k_r - \frac{c}{\omega}n\frac{\partial n}{\partial V_r}, \tag{8.f}$$

$$\frac{\partial H}{\partial k_\theta} = \frac{c^2}{\omega^2}k_\theta - \frac{c}{\omega}n\frac{\partial n}{\partial V_\theta}, \tag{8.g}$$

$$\frac{\partial H}{\partial k_\varphi} = \frac{c^2}{\omega^2}k_\varphi - \frac{c}{\omega}n\frac{\partial n}{\partial V_\varphi}, \tag{8.h}$$

$$\mathbf{k}\cdot\frac{\partial H}{\partial \mathbf{k}} = k_r\frac{\partial H}{\partial k_r} + k_\theta\frac{\partial H}{\partial k_\theta} + k_\varphi\frac{\partial H}{\partial k_\varphi}, \tag{8.i}$$

dove:

$$n' = n + f\frac{dn}{df} = n + \omega\frac{dn}{d\omega}, \tag{9}$$

$$V_r^2 + V_\theta^2 + V_\varphi^2 = \mathrm{Re}[n^2]. \tag{10}$$

L'ultimo passaggio è quello di sostituire al posto di $n$ il valore che esso assume nella formula di Appleton-Hartree

$$n^2 = 1 - \frac{X}{1 - jZ - \frac{Y_T^2}{2(1-X-jZ)} \pm \sqrt{\frac{Y_T^4}{4(1-X-jZ)^2} - Y_L^2}}, \tag{11}$$

dove $X$, $Y$ e $Z$ sono parametri adimensionali noti nella teoria magnetoionica e si riferiscono rispettivamente al quadrato del rapporto tra la frequenza di plasma e la frequenza dell'onda, al rapporto tra la frequenza ciclotronica e la frequenza dell'onda e al rapporto tra la frequenza di collisione e la frequenza dell'onda (Bianchi 1990, Davies 1965). In formule avremo

$$X = \frac{f_N^2}{f^2} = \frac{\omega_N^2}{\omega^2}, \tag{12a}$$

$$Y = \frac{f_{cicl}}{f} = \frac{\omega_{cicl}}{\omega}, \tag{12b}$$

$$Z = \frac{\nu}{f}, \tag{12c}$$

a seconda dell'impiego della frequenza *f*, o della pulsazione $\omega$. I pedici della Y indicano la componente trasversale T e longitudinale L del campo magnetico terrestre rispetto alla direzione del vettore d'onda **k**. Ovviamente, la frequenza di plasma dipende dalla densità elettronica N, la frequenza ciclotronica dal campo d'induzione magnetico terrestre **B** e la frequenza di collisione dalla densità della componente neutra del magnetoplasma minoritario.

## Bibliografia